\documentclass[a4paper,5p, authoryear,twoside,10pt]{elsarticle}

\usepackage{graphicx}
\usepackage{amsmath}
\usepackage{verbatim} 
\usepackage{lineno} 
\usepackage{multirow}

\journal{Icarus}

\newcommand*\patchAmsMathEnvironmentForLineno[1]{%
  \expandafter\let\csname old#1\expandafter\endcsname\csname #1\endcsname
  \expandafter\let\csname oldend#1\expandafter\endcsname\csname end#1\endcsname
  \renewenvironment{#1}%
     {\linenomath\csname old#1\endcsname}%
     {\csname oldend#1\endcsname\endlinenomath}}% 
\newcommand*\patchBothAmsMathEnvironmentsForLineno[1]{%
  \patchAmsMathEnvironmentForLineno{#1}%
  \patchAmsMathEnvironmentForLineno{#1*}}%
\AtBeginDocument{%
\patchBothAmsMathEnvironmentsForLineno{equation}%
\patchBothAmsMathEnvironmentsForLineno{align}%
\patchBothAmsMathEnvironmentsForLineno{flalign}%
\patchBothAmsMathEnvironmentsForLineno{alignat}%
\patchBothAmsMathEnvironmentsForLineno{gather}%
\patchBothAmsMathEnvironmentsForLineno{multline}%
}

\begin{document}
\begin{frontmatter}

\title{Influence of an inner core on the long-period \\forced librations of Mercury}
\author[rob]{Marie Yseboodt\corref{cor1}}
\ead{m.yseboodt@oma.be}
\cortext[cor1]{Corresponding author}
\author[rob]{Attilio Rivoldini}
\author[rob]{Tim Van Hoolst}
\author[can]{Mathieu Dumberry}
\address[rob]{Royal Observatory of Belgium, 3 Avenue Circulaire, 1180 Brussels, Belgium, +32 2 790 39 52}
\address[can]{Department of Physics, University of Alberta, Edmonton, T6G 2E1, Canada}

\begin{abstract}
The planetary perturbations on Mercury's orbit lead to long-period forced librations of 
Mercury's mantle. 
These librations have previously been studied for a planet with two layers: a mantle and 
a liquid core. Here, we calculate how the presence of a solid inner core in the liquid 
outer core influences the long-period forced librations. 
Mantle-inner core coupling affects the long-period libration dynamics mainly by changing 
the free libration: first, it lengthens the period of the free libration of the mantle, 
and second, it adds a second free libration, closely related to 
the free gravitational oscillation between the mantle and inner core. 
The two free librations have periods between 2.5 and 18 y depending on the internal structure. 
We show that large amplitude long-period librations of 10's of arcsec are generated 
when the period of a planetary forcing approaches one of the two free libration periods. 
These amplitudes are sufficiently large to be detectable by spacecraft 
measurements of the libration of Mercury. 
The amplitudes of the angular velocity of Mercury's mantle at planetary forcing periods 
are also amplified by the resonances, but remain
much smaller than the current precision of Earth-based radar 
observations unless the period is very close to a free libration period. 
The inclusion of mantle-inner core coupling in the 
rotation model does not significantly improve the fit to the radar observations.
This implies that it is not yet possible to determine the size of the inner core of 
Mercury on the basis of available observations of Mercury's rotation rate.
Future observations of the long-period librations may be used to constrain the interior 
structure of Mercury, including the size of its inner core. 
\end{abstract}

\end{frontmatter}
%\linenumbers 
\section*{Introduction}
Mercury has a peculiar rotation: three rotation periods correspond to two revolution periods. 
This spin-orbit resonance leads to interesting physical phenomena such as the 
longitudinal librations. The librations are caused by the non-spherical mass distribution of 
Mercury, on which the Sun exerts a gravitational torque. 
The difference between the orbital and the rotation periods leads to a varying torque
along the orbit since the orientation of the long axis of Mercury changes with respect 
to the direction to the Sun. Eccentricity further contributes to the variability by causing 
changes in the distance to the Sun and in the orbital speed.
The main libration has a period of 87.97 days, equal to Mercury's orbital (annual) period, 
and an amplitude of $38.5 \pm 1.6$ arcsec \citep{Marg12}. In addition, there are smaller 
amplitude librations at harmonic (semi-annual, ter-annual,...) frequencies. 
Those librations depend on the interior structure, notably the presence and the size of a liquid 
core inside the planet. By measuring the librations, we can infer knowledge about 
the interior structure. 
For example, by measuring Mercury's 88d libration amplitude, \citet{Marg07} 
concluded that Mercury has a large liquid core. 
Since a magnetic field has been detected by the Mariner 10 spacecraft 
\citep[e.g.][]{Ness75}, it is thought
that Mercury may have a solid inner core inside its liquid core.
\citet{Peal02} and \citet{Veas11} have investigated the consequence 
of the addition of a solid inner core on the rotation dynamics of Mercury. 
Recently, \citet{Vanh12} showed that, if the inner core is larger than about 1000 km, 
the difference on the 88d libration amplitude may be non-negligible, and of the same order as 
the present uncertainty, about 1.5 as. 

Another forced libration results from planetary perturbations.
The periodic force arising from the gravitational interaction of a planetary body with Mercury 
causes a perturbation of Mercury's orbital motion, changing its position relative 
to the Sun and thus altering the solar torque acting on its equatorial bulge.
This is an indirect effect of the planets on the rotation of Mercury.
These long-period forced librations induced by the planetary perturbations have been predicted 
by e.g. \citet{Dufe08}, \citet{Peal09} and \citet{Yseb10}.
They have periods commensurate with the orbital revolution of the planets
involved and are expected to have small amplitudes unless their period is close to 
the period of a free libration in which case a
near-resonant amplification can occur. In the absence of an inner core, there is only one such mode, 
the free mantle libration. This mode describes an oscillation of
the axis of minimum moment of inertia about the Mercury-Sun line 
at perihelion \citep{Peal05}. The period of the free mantle libration depends on the 
moments of inertia of Mercury, and is approximately 12 years. This is very close to 
Jupiter's perturbation on Mercury's orbit at 11.86y; \citet{Dufe08} and \citet{Peal09} 
have shown that a forced libration of 20 arcsec or more can be generated, the exact amplitude 
depending on the moments of inertia. Besides this 11.86y forced libration, at least 
4 other long-period forced librations have amplitudes larger than the arcsecond level.

The previous studies on the long-period forced librations assumed no mantle-inner core coupling.
Adding an inner core has two effects on the free libration: 
First, as shown by \citet{Peal02}, \citet{Veas11} and \citet{Vanh12}, it can 
lengthen the period of the free libration of the mantle
since the motion of the mantle is locked to that of the inner core for this mode.
As a result, interior models with a
large inner core may no longer have a free period close to Jupiter's orbital period. 
Second, the presence of the inner core adds a second free libration, closely 
approximated by the free gravitational oscillation between the mantle and inner core, 
and thus the possibility of additional resonances at other orbit perturbation frequencies.

In this study, we investigate how a non-spherical inner core 
coupled to the mantle and the outer core may influence the long-period forced librations. 
Since planetary perturbation periods may be close to the period of the two free modes, 
the long-period librations may be resonantly enhanced and
this may lead to a rotation angle that substantially differs from a model where 
there is no solid inner core. 
The presence of the solid inner core affects the rotation state of Mercury and may result 
in a signature that is detectable in the observations, in which case it must be taken
into account when analyzing the data.
In our rotation model, we also take into account the dissipation since this effect reduces 
the libration amplitudes and introduces phase lags. 
We include viscous and electromagnetic coupling at the core-mantle and inner-outer
core boundaries, as well as the effect of viscous deformation within the inner core. 
The signature of the parameters responsible for the dissipation on the libration is discussed.

In the theory section, we derive equations for the amplitude of the long-period 
forced librations of the mantle and of the inner core.
The equations are given for cases with and without dissipation.
We then numerically evaluate these libration equations on the basis of recent interior 
models of Mercury's \citep{Rivo09} (section 1.7). The results are compared for different 
interior models.
In the results section (section 2), we show that the amplitude of the long-period 
librations are of the order of a few arcsec, and much larger if the forcing period 
approaches the period of one of the free modes. 
In the last section (section 3), we compare predictions of libration models with and without 
mantle-inner core coupling with the Earth-based radar observations of the rotation rate 
of Mercury of \citet{Marg12} in order to determine whether the size of the inner 
core can be determined on the basis of the currently available radar data.

\section{Theory}

\subsection{Equations of motion for the mantle and the solid inner core} 

We assume an equatorial flattened bi-axial model of Mercury with 
a silicate shell composed of the mantle and the crust (we use the symbol $m$ for the shell), 
a fluid outer core ($oc$) and the solid inner part of the core ($ic$).
If the mantle and the inner core have a different rotation, their principal axes of 
inertia will be misaligned and there will be an effect of the gravitational and pressure 
coupling between these layers (see for example \citealt{Vanh12}).
The librational motion of the mantle and the solid inner core can be described by 
considering the change in angular momentum of these layers as a result of the external 
torque of the Sun and the internal torques. For the libration equation, we assume that 
the mantle and inner core are rigid solids as the effect of elastic deformations has been 
shown to be below the observational detection limit \citep{Vanh12}.
We then have the equations of motion:
\begin{align}
\ddot{\psi}_m &=
- \frac{G M_S}{C_m\,r^3\,n^2} \,K_m \, \sin 2 \left(\psi_m - \varpi - f \right) 
- \frac{K}{C_m} \sin 2 \left(\psi_m -\psi_{ic} \right) \; ,
\label{eq_psm} \\
\ddot{\psi}_{ic} &=
- \frac{G M_S}{C_{ic}\,r^3\,n^2} \,K_{ic} \, \sin 2 \left(\psi_{ic} - \varpi - f \right) 
+ \frac{K}{C_{ic}} \sin 2 \left(\psi_m -\psi_{ic} \right) \; .
\label{eq_pss}
\end{align}
The rotation angle of the mantle $\psi_m$ describes the orientation of the axis of 
minimum moment of inertia of the mantle $A_m$ relative to 
the intersection between the ecliptic and the orbital plane at J2000. 
Similarly, $\psi_{ic}$ is the rotation angle of the inner core.
$f$ is the true anomaly, 
$\varpi$ the longitude of the pericenter,
$r$ the distance between the mass centers of Mercury and the Sun,
$A_{ic}\,<\,B_{ic}\,<\,C_{ic}$ are the principal moments of inertia of 
the inner core and
$A_m\,<\,B_m\,<\,C_m$ the mantle moments of inertia. 
$n$ is the mean motion of Mercury, $M_S$ the mass of the Sun and $G$ the gravitational constant. 
The factors $K_m$ and $K_{ic}$ describe the strengths of 
the gravitational and pressure torques on the mantle and inner core, respectively 
\citep{Vanh12,Dumb13}.
These factors are defined by 
$K_m=\frac{3} {2} \, n^2 \left( B_m-A_m + B_{oc,t}-A_{oc,t} \right)$
and
\newline $K_{ic}=\frac{3} {2} \, n^2 \left( B_{ic}-A_{ic}+B_{oc,b}-A_{oc,b}\right)$, where
$A_{oc,b}$ and $B_{oc,b}$ are the principal moments of inertia of the 
bottom part of the fluid core (a layer between the inner core-outer core boundary (ICB)
and the smallest sphere that can be included in the fluid core, see Fig.~2 of \citealt{Vanh08})
while $A_{oc,t}$ and $B_{oc,t}$ are related to the rest of the fluid core.
The terms proportional to $B_m-A_m$ and $B_{ic}-A_{ic}$ capture the solar 
gravitational torques on the mantle and inner core, respectively. The additional 
terms arise from the pressure torques on the boundaries
between the outer core and mantle and between the inner core and outer core.
It can be shown that in the limit of no inner core, the expression of $K_m$ reduces to 
$(3/2) \, n^2 (B-A)$, and we retrieve the classical equation of a planet with two layers. 
$K$ is the gravitational-pressure coupling constant between the mantle and the inner core. 
If the inner and outer parts of the core have uniform density $\rho_j$ 
($j=oc$ for the fluid outer core, $j=ic$ for the solid inner core and $j=m$ for the 
silicate shell), $K$ is defined by \citep[e.g.][]{Veas11}
\begin{equation}
K = \frac{4 \pi G}{5}\left(1-\frac{\rho_{oc}}{\rho_{ic}}\right) C_{ic}\, 
\beta_{icb}\left[ (\rho_{oc}-\rho_m)\,\beta_{cmb}+\rho_m\beta_m\right] \, ,
\label{eq_K}
\end{equation}
where $\beta_m$, $\beta_{cmb}$ and $\beta_{icb}$ are the geometrical equatorial flattenings
at the top of the mantle, core-mantle boundary (CMB) and ICB, respectively. 
If an ellipsoidal surface of constant density at a given radius has its three principal semi-axes 
defined by $a>b>c$, the geometrical flattening in the equatorial plane is defined by 
$\beta=({a-b})/{a}$. 
For the computation of the longitudinal librations, the polar flattening may be neglected.
When radial density variations in both the fluid and solid cores are taken into account, 
the expression for $K$ is more complicated and given in \citet{Dumb13}. 
Since the effect of the small obliquity on the longitudinal librations is below the observational
detection limit, the obliquity of Mercury is assumed to be $0$ (its true value is $0.034^\circ$, 
\citealt{Marg12}) so that Mercury's equator and orbit are the same plane. 

Previous studies of the effect of the inner core on Mercury's rotation focused on the 
amplitude of the 88d librations and considered a Kepler orbit, in which
the orbital elements are constant with time, except for the true and the mean anomalies.
In order to derive differential equations that use a small angle,
the rotation angle $\psi_m$ is usually related to the mantle libration angle $\gamma_m$
with the relation $\psi_m = \gamma_m + 1.5 \, M$, where $M = n(t-t_0)$ is the mean anomaly 
of the orbital motion and $t_0$ is a chosen initial time. 
A similar expression is used to relate $\psi_{ic}$ and $\gamma_{ic}$, the libration angles 
of the inner core.

When long-period librations are considered, the longitude of the pericenter $\varpi$ is no longer 
constant with time while the true anomaly $f(t)$ and the distance between Mercury and the Sun 
$r(t)$ have a quasiperiodic evolution because of the planetary perturbations that 
affect the orbit of Mercury. 
Following \citet{Yseb10} we define the mantle libration angle 
$\gamma_m$ and an inner core libration angle $\gamma_{ic}$ by 
\begin{align}
\psi_m&= \gamma_m + 1.5 \, M + \varpi = \gamma_m + \lambda
\label{eq_defgam}\, ,\\
\psi_{ic}&=\gamma_{ic} + 1.5 \, M + \varpi = \gamma_{ic} + \lambda
\label{eq_defgas}\, ,
\end{align}
where the angle $\lambda$ is defined by $1.5 \, M + \varpi$. 
All these angles are functions of time.
The angle $\lambda$ can be written as a sum of the mean rotation and 
a frequency decomposition
$\lambda(t)= 1.5 \,n\, t + 1.5 \, M(t_0) + \varpi(t_0) + \dot \varpi 
+ \sum_{i} \lambda_i \cos (\omega_i \, t + \phi_{\lambda_i})$
where $\lambda_i$, $\omega_i$ and $\phi_{\lambda_i}$ are the amplitude, angular frequency 
and phase of the different planetary perturbations on Mercury's orbit. 
The origin of the angles $\psi_m$ and $\psi_{ic}$ is defined by a fixed line with respect 
to an inertial frame, while the origin of the $\gamma_m$ and $\gamma_{ic}$ angles 
is moving because of the planetary perturbations (see figure 1 in \citealt{Yseb10}).
Only $\psi_m$ can be observed although the libration angle of the mantle $\gamma_m$ 
is more convenient mathematically because it is a small quantity.

To compute the long-period forced librations, the differential equations (\ref{eq_psm}) 
and (\ref{eq_pss}) 
are expanded as two functions of the mean anomaly and of the eccentricity.
Since we study the long-period librations, we discard the terms containing 
the mean anomaly of Mercury in the right hand side of the equations. However terms 
containing the second time derivative of the mean anomaly and the longitude of the 
pericenter ($\ddot \lambda(t)$) remain in the left hand side of the equations \citep{Yseb10}.
The angles $\gamma_m$ and $\gamma_{ic}$ are usually small, so that 
the small angle approximation can also be used. We get two linearized equations for 
$\gamma_m$ and $\gamma_{ic}$ that contain the planetary perturbations on the orbit of Mercury
\begin{align}
\ddot{\gamma}_m - \sum_{i} \omega_i^2 \lambda_i \cos (\omega_i t + \phi_{\lambda_i}) &=
- \, \omega_m^2 \, \gamma_m - \frac{2 K}{C_m} \left( \gamma_m -\gamma_{ic} \right) \, ,
\label{eq_gamF}\\
\ddot{\gamma}_{ic} - \sum_{i} \omega_i^2 \lambda_i \cos (\omega_i t + \phi_{\lambda_i}) &=
- \, \omega_{ic}^2 \, \gamma_{ic} + \frac{2 K}{C_{ic}} \left( \gamma_m -\gamma_{ic} \right)\, ,
\label{eq_gasF}
\end{align}
where the frequencies $\omega_m$ and $\omega_{ic}$ are given by 
\begin{eqnarray}
\omega_m^2 &=& 2 \, f_2 \,\frac{ K_m }{C_m}\, ,
\label{eq_omm2}\\
\omega_{ic}^2 &=& 2 \, f_2 \,\frac{ K_{ic}}{C_{ic}}\, ,
\label{eq_oms}
\end{eqnarray}
and where $f_2$ is a function of the eccentricity ($f_2(e)=\frac{7}{2}e-\frac{123}{16}e^3+\cdots\,$).

In the linearized equations (\ref{eq_gamF}) and (\ref{eq_gasF}), 
the planetary perturbations are not included as a torque acting on the 
mantle and the inner core (this direct effect is negligible because of the large distance 
between Mercury and the other planets) but appear as an indirect effect, 
through the definition of the $\gamma$ angles. 
This additional term, coming from the time derivative of the angle $\psi_m$,
acts like a ''forcing'' in the differential equations.
In the limit of no inner core, $K=0$ and Eq.~(\ref{eq_omm2}) becomes 
the usual expression of the free mantle libration frequency 
\begin{equation}
\omega_{m\; \textrm{no ic}}^{2} = 3 \,n^2 \,f_2 \,(B-A)/C_m \, \cdot
\label{eq_omm}
\end{equation}

\subsection{Planetary perturbations from the ephemerides}
Using the JPL DE405/DE406 ephemerides \citep{Stan98} over a period of about 1000 years 
before and after J2000 and performing a frequency analysis, 
we get the frequencies $\omega_i$, amplitudes $\lambda_i$ and phases $\phi_{\lambda_i}$ of 
the main planetary perturbations on Mercury's orbit (see Table 1).
We use the frequency mapping FAMOUS software (ftp://ftp.obs-nice.fr/pub/mignard/Famous).
\begin{center}
\begin{tabular}{|c c|c c|c|}
\hline
Period & Forcing & $\lambda_i$ & $\phi_{\lambda_i}$ & $\psi_{m\,i}$ \\
$2 \pi/w_i$ (y) & argument& (as)& (deg) & (as) \\
\hline
3.955  & Jupiter ($3\lambda_J$) & 0.45 & 35& 0.1\\
5.663  & Venus ($2\lambda_M - 5\lambda_V$) & 12.7	& 87	& 3.9\\
5.932  & Jupiter ($2\lambda_J$)	& 4.31	& 4	& 1.5\\
6.574  & Earth ($\lambda_M - 4\lambda_E$) & 1.37 & 332	& 0.6\\
11.861 & Jupiter ($\lambda_J$)		& 1.40	& 171 &	38.4 \\
14.727 & Saturn ($2\lambda_S$)		& 0.52	& 36&	 1.4\\
\hline
1.110 & ($\lambda_M - 2\lambda_V$) & 7.59 & 235& 0.1\\
0.555 & (2 $\lambda_M - 4\lambda_V$) & 1.01 & 17& 0.0\\
0.292 & (2 $\lambda_M - 3\lambda_V$) & 0.99 & 305& 0.0\\
0.465 & ($\lambda_M - 2\lambda_E$) & 0.88 & 38& 0.0\\
0.251 &  & 0.79 & 172& 0.0\\ 
0.396 & ($\lambda_M - \lambda_V$) & 0.72 & 160& 0.0\\
0.615 & ($\lambda_V$) & 0.65 & 16 &0.0\\
0.198 &  & 0.64 & 231 &0.0\\ 
0.241 & & 0.62 & 86& 0.0\\ 
1.380 & ($\lambda_M - 3\lambda_V$) & 0.40 & 222 &0.0\\
\hline
\end{tabular}
 
Table 1: Planetary perturbations of Mercury's orbit and 
frequency decomposition of the $\lambda = 1.5 \,M + \varpi$ angle. 
The last column shows the amplitude of the librations (rotation angle $\psi_{m}$, see 
Eq.~(\ref{eq_psimnoPl})) if $(B-A)/C_m = 2.18 \times 10^{-4}$ \citep{Marg12} and if there 
is no mantle-inner core coupling.
The six perturbation frequencies that may be resonantly amplified by free librations are 
listed above the horizontal line.
\end{center}
Table 1 is an updated version of Table 1 in \citet{Yseb10}. 
Some typos have been corrected and some values are slightly altered because we use a 
different time interval for the ephemerides. 
Using another ephemerides will also slightly change these values.
In the second column we give the source of the planetary perturbation (forcing argument).
Note that the origin of some of these perturbations (mostly the periods smaller than 100 days) 
cannot be identified unambiguously and are left blank; these do not match exactly a known 
orbital frequency and may represent the combined effect from many closely separated frequencies. 
The main planetary perturbations have periods that range from approximately twenty days to 
fifteen years and the amplitudes of the $\lambda$ angle are as high as 12.7 as.

\subsection{The eigenmodes}
The frequencies of the eigenmodes are found by solving for the roots of the characteristic equation 
$\omega^4 - 2\, \omega^2 \, \omega_a^2 + \left( \omega_{ic}^2\,\alpha_m^2 + \omega_m^2 \,\alpha_{ic}^2 
+ \omega_m^2 \,\omega_{ic}^2 \right) = 0$
of Eqs.~(\ref{eq_gamF}) and (\ref{eq_gasF})
\citep{Dumb11,Vanh12}
\begin{align}
\omega_{1,2}^2 &= \omega_a^2 
\mp \sqrt{\omega_a^4 - \,\left(\omega_{ic}^2 \, \omega_m^2 + \omega_m^2\, \alpha_{ic}^2 + 
\omega_{ic}^2 \, \alpha_m^2 \right)}\, ,
\label{eq_om1} 
\end{align}
where 
\begin{equation}
\omega_a^2 =\frac{ \omega_g^2+\omega_m^2+\omega_{ic}^2}{2}, \quad
\omega_g^2 = \frac{2 K (C_m + C_{ic})}{C_{ic} \, C_m} \, , 
\end{equation}
\begin{equation}
\alpha_m^2=\frac{2K}{C_m}, \quad 
\alpha_{ic}^2=\frac{2K}{C_{ic}} \, \cdot
\label{eq_omg}
\end{equation}
Note that the frequencies of the free modes are independent of the planetary perturbations.

Free modes are natural solutions of the libration equations and, if excited by some process, 
may have arbitrary large amplitudes. Dissipation is undoubtedly present (and will be 
discussed in section 1.5) and will attenuate the amplitude of the free librations in time
on a timescale much smaller than the age of the Solar System \citep{Peal05}. 
So if free modes are part of Mercury's librations, they require a recent or 
on-going excitation. 
The most recent analysis of Mercury's spin rate observations suggests that a libration 
model that includes a free mantle libration does not significantly improve the fit to observations 
\citep{Marg12}. However, the spin rate variations associated with a decadal free 
librations of a few 10's of arcsec are much smaller than those caused by the much more 
rapid 88d libration. Thus, it is difficult to rule out conclusively the presence of free 
librations on the basis of spin rate data alone. Robust conclusions on the amplitude (and 
period) of the free librations must await future observations, especially observations of the 
long term changes in the angle of libration.

\subsection{The amplitude of the undamped long-period forced librations}

Without mantle-inner core coupling and without dissipation but taking into account the
planetary perturbations on the orbit of Mercury, 
the analytical solution of the linearized differential equation~(\ref{eq_gamF}) for the 
long-period forced libration of the mantle is \citep{Yseb10}
\begin{equation}
\gamma_m(t)= \sum_{i}{\lambda_i \frac{\omega_i^2}{\omega_m^2-\omega_i^2} 
\cos (\omega_i \, t + \phi_{\lambda_i})}\, ,
\label{eq_gamannoPl}
\end{equation}
or the long-period forced librations for the rotation angle $\psi_m$
\begin{equation}
\psi_m(t)= \sum_{i}{\lambda_i \frac{\omega_m^2}{\omega_m^2-\omega_i^2} 
\cos (\omega_i \, t + \phi_{\lambda_i})}\, \cdot
\label{eq_psimnoPl}
\end{equation}
The long-period forced librations have angular frequencies $\omega_i$ equivalent to the 
orbital frequencies.

The expression for the forced librations of the inner core $\gamma_{ic}(t)$
including planetary perturbations but neglecting any coupling between the mantle 
and the inner core is very similar to the expression $\gamma_m(t)$, the
$\omega_m$ frequency being replaced by $\omega_{ic}$.
The inner core free libration (see Eq.~(\ref{eq_oms})) has a period of about $50-60$y.

If the effect of mantle-inner core coupling on the libration and the planetary perturbations 
on the orbit are included, the forced part of the solution for $\gamma_m$ and $\gamma_{ic}$ 
of Eqs.~(\ref{eq_gamF}) and (\ref{eq_gasF}) are
\begin{align}
\gamma_m(t) &= 
 \sum_{i}{\lambda_i \, \frac{\omega_g^2 + \omega_{ic}^2 -\omega_i^2}
{\left(\omega_1^2-\omega_i^2\right) \left(\omega_2^2 -\omega_i^2\right)} 
\omega_i^2 \cos (\omega_i t + \phi_{\lambda_i})} \, ,
\label{eq_gamAn}\\
\gamma_{ic}(t) &= \sum_{i}{\lambda_i \, \frac{ \omega_g^2 + \omega_m^2 -\omega_i^2}
{\left(\omega_1^2-\omega_i^2\right) \left(\omega_2^2 -\omega_i^2\right)} 
\omega_i^2 \cos (\omega_i t + \phi_{\lambda_i})} \, \cdot
\label{eq_gasAn}
\end{align}
By using Eq. \ref{eq_defgam}, the rotation angle $\psi_m$ of the mantle can be written as
\begin{align}
\psi_m(t) &= \sum_{i}{\lambda_i \, \frac{\alpha_m^2 \, \omega_{ic}^2+ \omega_m^2 
( \alpha_{ic}^2 +\omega_{ic}^2 -\omega_i^2 ) }
{\left(\omega_1^2-\omega_i^2\right) \left(\omega_2^2 -\omega_i^2\right)} 
\cos (\omega_i t + \phi_{\lambda_i})} \, ,
\label{eq_psimAn}\\
 &= \sum_{i}{\lambda_i \, \left( \frac{N} {\omega_1^2-\omega_i^2} 
+ \frac{\omega_m^2-N} {\omega_2^2-\omega_i^2}\right)
\cos (\omega_i t + \phi_{\lambda_i})} \, ,
\label{eq_psimAnDec}
\end{align}
where
\begin{equation}
N = \frac{\omega_m^2}{2} + \frac{\omega_m^2 (\alpha_{ic}^2 -\alpha_m^2 
- \omega_m^2 + \omega_{ic}^2) + 2 \,\alpha_m^2 \, \omega_{ic}^2 }
{4 \, \sqrt{\omega_a^4 - \,\left(\omega_{ic}^2 \, \omega_m^2 + \omega_m^2\, \alpha_{ic}^2 
+ \omega_{ic}^2 \, \alpha_m^2 \right)} } \, \cdot
\label{eq_N}
\end{equation}
By comparing Eqs.~(\ref{eq_psimnoPl}) and (\ref{eq_psimAn}), we notice
the second eigenfrequency in the denominator and a different numerator of the libration amplitude, 
while the dependence on the planetary perturbation $\lambda_i$ and the 
temporal behavior with the frequency and the phase do not change.
In Eq.~(\ref{eq_psimAnDec}), the two terms inside the parenthesis represent the resonance factors 
associated with each of the two free librations.
In this expression of the libration amplitude for the rotation angle $\psi_m$, 
the forcing frequency dependence is in the denominators only.
Additionally, for an equivalent offset to the resonance frequency,
a larger numerator of the resonance factor indicates a stronger resonance effect.
The resonance with the first eigenfrequency usually has a larger effect than 
the second eigenfrequency.

\subsection{The damped long-period forced librations}
\label{sec_diss}
Dissipation at the CMB from viscous or electromagnetic (EM) coupling 
between the mantle and the fluid core is proportional to their differential rotation rate. 
The effect of dissipation on libration is included by adding a term $-k_m(\dot{\gamma}_m -\dot{\gamma}_{oc})$ 
on the right-hand side 
of Eq.~(\ref{eq_gamF}), where $k_m$ is a damping coefficient \citep[e.g.][]{Peal05} and ${\gamma}_{oc}$ 
is the libration angle of the fluid core (and $\dot{\gamma}_{oc}$ its time-derivative). 
Likewise, dissipation at the ICB is introduced by including a term $-k_{ic} (\dot{\gamma}_{ic} -\dot{\gamma}_{oc})$ 
on the right-hand side of Eq.~(\ref{eq_gasF}), where $k_{ic}$ is a damping coefficient at the ICB. 
Because of the dissipative interaction of the outer core with the mantle and the inner core, 
the outer core will also librate and
a third equation, the evolution of the libration of the fluid core, is required to solve the system. 
We also add the dissipation from viscous deformation within the inner core. We follow 
\citet{Koni13} and assume an inner core with a simple Newtonian rheology 
with viscous deformation taking place over a characteristic e-folding time of $\tau$. 
This adds a term $-\tau^{-1}(\dot{\gamma}_{ic} -\dot{\gamma}_{m})$ on the right-hand side of 
Eq.~(\ref{eq_gasF}).

It is more convenient to re-write the forcing term in Eqs.~(\ref{eq_gamF}-\ref{eq_gasF}) as
\begin{equation}
\omega_i^2 \lambda_i \cos(\omega_i t + \phi_{\lambda_i}) 
= \omega_i^2 \left( f_i \, e^{-I \omega_i t} + f_i^* \, e^{I \omega_i t} \right) \, ,
\end{equation}
where $f_i^*$ is the complex conjugate of $f_i$ and $I=\sqrt{-1}$. 
The real and imaginary parts of $f_i$ are related to $\lambda_i$ and $\phi_{\lambda_i}$ by
\begin{eqnarray}
\mbox{Re}[f_i] &=& \frac{1}{2} \lambda_i \cos \phi_{\lambda_i} \,,\\
\mbox{Im}[f_i] &=& -\frac{1}{2} \lambda_i \sin \phi_{\lambda_i} \, \cdot
\end{eqnarray} 
Since equations (\ref{eq_gamF}-\ref{eq_gasF}) are linear in the libration angles $\gamma_m$ 
and $\gamma_{ic}$, we can find the solutions to the forcing $f_i$ (at frequency $\omega_i$) 
and that to the forcing $f_i^*$ (at frequency -$\omega_i$) individually and add up their solution. 
Let us do the $f_i$ part. The equations of motion for the mantle, fluid outer core and solid inner core are now
\begin{align}
\ddot{\gamma}_m + \, \omega_m^2 \gamma_m + \alpha_m^2 \left( \gamma_m -\gamma_{ic} \right) 
+ k_m (\dot{\gamma}_m - \dot{\gamma}_{oc}) &= \omega_i^2 f_i\,,
\label{eq_gamDI}\\
\ddot{\gamma}_{oc} -\frac{C_m}{C_{oc}} \,k_m (\dot{\gamma}_m - \dot{\gamma}_{oc}) 
- \frac{C_{ic}}{C_{oc}} \,k_{ic} \,(\dot{\gamma}_{ic} - \dot{\gamma}_{oc}) & = 0 \,,
\label{eq_gafDI}\\
\ddot{\gamma}_{ic} + \omega_{ic}^2 \gamma_{ic} - \alpha_{ic}^2 \left( \gamma_m -\gamma_{ic} \right) 
+ k_{ic} (\dot{\gamma}_{ic} - \dot{\gamma}_{oc}) + \frac{\dot{\gamma_{ic}} - \dot{\gamma}_m }{\tau} 
&= \omega_i^2 f_i\,\cdot
\label{eq_gasDI}
\end{align}

Assuming periodic solutions of the form 
\begin{eqnarray}
\gamma_m(t) & = &\tilde{\gamma}_{m} \, e^{-I \omega_i t}\,,\\
\gamma_{oc}(t) & = &\tilde{\gamma}_{oc} \, e^{-I \omega_i t}\,,\\
\gamma_{ic}(t) & = &\tilde{\gamma}_{ic} \, e^{-I \omega_i t}\,,
\end{eqnarray}
where $\tilde{\gamma}_{m}$, $\tilde{\gamma}_{oc}$, and $\tilde{\gamma}_{ic}$ are complex amplitudes, 
the system of equations (\ref{eq_gamDI}-\ref{eq_gasDI}) can be written as 
\begin{equation}
\mathsf{A} \cdot {\bf x} = {\bf f}\,,
\label{eq_syst}
\end{equation}
where the matrix $\mathsf{A}$ contains the model parameters and vectors ${\bf x}$ and ${\bf f}$ 
contain the forced libration amplitudes and forcing, respectively, 
\begin{equation}
{\bf x} = \left[
\begin{array}{c}
\tilde{\gamma}_{m} \\ \tilde{\gamma}_{oc} \\ \tilde{\gamma}_{ic}
\end{array}
\right] \, ,
\hspace*{1cm}
{\bf f} = \left[
\begin{array}{c}
\omega_i^2 f_i \\ 0\\ \omega_i^2 f_i
\end{array}
\right] \, \cdot
\label{eq_xf}
\end{equation}
The amplitude of libration (${\bf x}$) at a given forcing frequency can be found by solving 
Eq.~(\ref{eq_syst}). The amplitude and phase of each of the mantle, inner core and fluid core 
libration can then be retrieved from the real and imaginary parts of the solution vector 
${\bf x}$. For instance, the amplitude of the mantle libration is 
$\sqrt{\mbox{Re}[\tilde{\gamma}_{m}]^2 + \mbox{Im}[\tilde{\gamma}_{m}]^2}$. 
To this, we must add the solution from the $f_i^*$ part of the forcing: the solution is ${\bf x}^*$. 
So the total mantle amplitude is $2\sqrt{\mbox{Re}[\tilde{\gamma}_{m}]^2 + \mbox{Im}[\tilde{\gamma}_{m}]^2}$.
If the phase of $\gamma_m(t)$ is defined as $\phi_{\lambda_i} + \phi^R_i$ (R like resonant), 
then the absolute value of the amplitude of the rotation angle $\psi$ at frequency $\omega_i$ is 
\begin{equation}
\tilde{\psi}_{m} = \sqrt{\tilde{\gamma}_m^2 + \lambda_i^2+ 2 \tilde{\gamma}_m\, \lambda_i \cos{\phi^R_i}}\,,
\label{eq_psigam}
\end{equation}
while the phase of the rotation angle $\psi_m$ is given by
\begin{equation}
\tan \phi_{\psi_i} = \frac{\tilde{\gamma}_{m} \sin{(\phi_{\lambda_i} + \phi^R_i)}+ \lambda_i \sin{\phi_{\lambda_i}}}
{\tilde{\gamma}_{m} \cos{(\phi_{\lambda_i} + \phi^R_i)}+ \lambda_i \cos{\phi_{\lambda_i}}}\, \cdot
\label{eq_psiphase}
\end{equation}
The analytical expressions of $\tilde{\gamma}_{m}$, $\tilde{\psi}_{m}$ and its phase 
as a function of the different frequencies $\alpha_m$, $\alpha_{ic}$, $\omega_m$ and 
$\omega_{ic}$ can be derived, they are too long to be printed here.

The inclusion of dissipation alters the amplitude of the long-period forced librations given in 
Eqs.~(\ref{eq_gamAn}-\ref{eq_psimAn}) and adds a phase lag between the planetary forcing and 
the response of the mantle, fluid core and inner core. The eigenfrequencies 
$\omega_1$ and $\omega_2$ are now complex. 
For small damping parameters ($k_m$ and $k_{ic}$ smaller than $10^{-3}$/y, 
$\tau >$ 100 years), 
the change in the real part of the eigenfrequencies is negligible. 
The introduction of dissipation through $k_m$ and $k_{ic}$ results in a finite amplitude 
of libration when the forcing period is equal to one of the free libration periods. The maximal amplitude 
of $\tilde{\psi}_{m}$ at the resonant frequency $\omega_i\approx \omega_1$ or $\omega_2$ in the limit of 
a rigid inner core is approximately equal to 
\begin{equation}
\tilde{\psi}_{m\, \textrm{MAX}} \approx \frac{\lambda_i}{\omega_i} \, \frac{\alpha_m^2 \, \omega_{ic}^2 + 
\alpha_{ic}^2 \,\omega_m^2 
- \omega_m^2\, \omega_i^2 + \omega_m^2\,\omega_{ic}^2}
{ k_m \left(\alpha_{ic}^2-\omega_i^2+\omega_{ic}^2\right)+ k_{ic} \left(\alpha_m^2-
\omega_i^2+\omega_m^2\right)} \, \cdot
\label{eq_psimimax}
\end{equation}

The amplitudes of the damping parameters $k_m$ and $k_{ic}$ depend on the nature of the coupling.
For viscous coupling, using a molecular value of the kinematic viscosity of Mercury's fluid core of the 
order of $10^{-6}$ m$^2$/s \citep[e.g.,][]{deWi98}, $k_m$ and $k_{ic}$ are both of the 
order of $10^{-5}$/y \citep{Peal05}. This validates the small damping approximation that led to 
expression (\ref{eq_psimimax}). 

Electromagnetic coupling at the CMB depends on the electrical conductivity of the lower mantle. If a conducting 
layer of thickness $\Delta$ and conductivity $\sigma_m$ is present at the base of Mercury's mantle, then 
$k_m^{\textrm{EM}}$ would be given by \citep[e.g.,][]{Buff98}
\begin{equation}
k_m^{\textrm{EM}} = \frac{\sigma_m \,\Delta}{C_m}\, r_{cmb}^4 \,\mathcal{I}(B_r)_{\textrm{CMB}}\, ,
\end{equation}
where $\mathcal{I}(B_r)_{\textrm{CMB}}$ is a factor that depends on the geometry of the 
radial magnetic field at the CMB and $r_{cmb}$ is the CMB radius. 
For a simple dipole magnetic field of RMS amplitude $B_r^d$, $\mathcal{I}(B_r)= 64\pi \, (B_r^d)^2\,/15$
\citep{Koni13}. Using a dipole field amplitude of 200 nT, of the same order as that 
observed by MESSENGER \citep{Ande11} and a mantle conductivity $\sigma_m=0.1$ S/m 
\citep[e.g.][]{Cons07,Verh09}, then even if $\Delta$ is equal to the whole 
mantle thickness, $k_m^{\textrm{EM}}$ is several orders of magnitude smaller than the viscous estimate above. 
If a layer of highly conducting material is present at the base of the mantle, as is inferred for Earth from 
nutation observations \citep[e.g.,][]{Buff92} or at the top of the core of
Mercury as it has been proposed by \citet{Smit12}, then $k_m^{\textrm{EM}}$ can be larger. 
However, even if we use the terrestrial values of $\Delta=200$ m and $\sigma_m=10^6$ S/m, because the 
magnetic field of Mercury is weak, electromagnetic coupling remains weaker than viscous coupling. 
Electromagnetic coupling at the CMB can be neglected.

Electromagnetic coupling can be larger at the ICB because both the inner core and fluid core are good conductors. 
If we assume the same electrical conductivity $\sigma$ on both the solid and fluid sides of the ICB, 
the EM coupling constant $k_{ic}^{\textrm{EM}}$ can be written as \citep[e.g.,][]{Buff98}
\begin{equation}
k_{ic}^{\textrm{EM}} = \frac{(1+I)}{4 \sqrt{ \omega_i}} \frac{1}{C_{ic}}\sqrt{\frac{2 \sigma}{ \mu}} 
\,r_{icb}^4 \,\mathcal{I}(B_r)_{\textrm{ICB}} \, ,
\end{equation}
where $\mu$ is magnetic permeability. The EM coupling parameter
$k_{ic}^{\textrm{EM}}$ is complex and frequency dependent. If we use a nominal
ICB magnetic field of the same order as that at the CMB, and a 5y periodic
forcing, then $k_{ic}^{\textrm{EM}}$ is approximately $10^{-5}$/y for
$\sigma=10^6$ S/m \citep{Deng13}, of the same order as for viscous coupling. It is likely that
the field at the ICB is larger than at the CMB, and if we use a factor 10
increase, then $k_{ic}^{\textrm{EM}} \approx 10^{-3}$/y. Although EM coupling
would dominate viscous coupling in this "strong field" regime, the amplitude of
$k_{ic}^{\textrm{EM}} $ remains sufficiently small as to not lead to large
variations in the period of the free librations calculated in the absence of
dissipation \citep[e.g.][]{Dumb11}. In all the calculations that we 
report in the results section, we use $k_m=k_{ic}=10^{-5}$/y, unless otherwise
noted.
 
Viscous dissipation within the inner core is difficult to evaluate because we do
not know the viscosity of the inner core and hence, the characteristic timescale
$\tau$. However, if $\tau$ is shorter than $10^5$y, then dissipation of the
libration energy through viscous inner core deformation should dominate that
from coupling at the CMB and ICB. Given that the temperature inside the inner
core is likely not too far from the melting point, it is conceivable that
$\tau$ may be as short as 10y. In the results section, we will report on the
effect of using different values of $\tau$ in our calculations.

\subsection{Harmonics of the 88d libration}
\label{sec88}
 
In order to compare the model to the data, we need a rotation model
that includes both the short-period and long-period forced librations.
We use all harmonics of the 88d forced librations up to the 6th harmonic.
\begin{equation}
\gamma_{m} = \sum_i^7{g_i \,\sin{(i\, n \,(t-t_\mathrm{pericenter}))}}\, \cdot
\end{equation}
The amplitude $g_i$ of the forced librations at the frequency $i\,n$ can be expressed as 
\begin{align}
g_i & = (G_{2\,0\,1-i}-G_{2\,0 \, 1+i}) \, . \nonumber \\
&\frac{ i^2 n^2 C_m C_{ic}\,\omega_m^2 - C_m C_{ic}\, \omega_m^2 \omega_{ic}^2
- 2 K C_m \,\omega_m^2 - 2 K C_{ic}\, \omega_{ic}^2}
{2 f_2 \, C_m C_{ic} \left(i^2 n^2-\omega_1^2\right) \left(i^2 n^2-\omega_2^2\right)} \, ,
\label{eq_annualharm}
\end{align}
where the $G_{j k l}$ are the eccentricity functions of Kaula (see Table 2 or \citet{Kaul66} for example).
\begin{center}
\begin{tabular}{|c | c|} 
\hline
$i$ & $G_{2\,0\,1-i}-G_{2\,0 \, 1+i}$\\
\hline
1 & $1 - 11 e^2 + \frac{959 \, e^4}{48} - \frac{3641 \, e^6}{288} $\\
2 & $-\frac{e}{2} - \frac{421 \, e^3}{24} + \frac{32515 \, e^5}{768}$\\
3 & $-\frac{533 \, e^4}{16} + \frac{13827 \, e^6}{160}$\\
4 & $\frac{e^3}{48} - \frac{57073 \, e^5}{960}$\\
5 & $\frac{e^4}{24} - \frac{18337 \, e^6}{180}$\\
6 & $\frac{81 \, e^5}{1280}$\\
7 & $\frac{4 \, e^6}{45}$\\
\hline
\end{tabular}

Table 2: Eccentricity functions of the annual period and its first six harmonics, 
up to degree 6 in eccentricity. 
\end{center}
Earth-based radar observations of Mercury yield estimates of the rotation rate variations.
Because of the time derivative of the rotation rate that emphasizes the weight of 
the short period terms with respect to the rotation angle, 
the number of harmonics needed to precisely describe the 
temporal behavior of the rotation rate is larger than the numbers of harmonics of the libration angle model.
The annual and the semi-annual amplitudes have been given in \citet{Vanh12}.
The ter-annual and the other small period libration amplitudes have been derived here
using the same method. However, the effect of the inner core on the ter-annual and the 
other very short period is very small (smaller than 0.1 as for the libration angle or 
smaller than 1 as/y for the rotation rate).

If the effect of the coupling of the inner core and the mantle is neglected then 
Eq.~(\ref{eq_annualharm}) is reduced to the expression 
\begin{equation}
g_i = \frac{(G_{2\,0\,1-i}-G_{2\,0 \, 1+i})}{i^2} \, \frac{3 \,(B-A)}{2 \,C_m- 6 f_2 (B-A)} \, \cdot
\label{eq_g88}
\end{equation}

\subsection{Interior model for Mercury and flattening of the layers}

The amplitude of the long-period forced librations depends on the frequencies of the
free modes $\omega_{1,2}$ as well as on $\omega_g$, $\omega_m$, $\omega_{ic}$,
$\alpha_m$ and $ \alpha_{ic}$. In turn, these quantities depend on the
parameters $K$, $K_m$ and $K_{ic}$, themselves being function of the moments of
inertia of the mantle, fluid outer core and solid inner core. To determine these quantities, we first construct
hydrostatic spherical interior models for Mercury with the method presented in \citet{Rivo09} 
and an updated model for the core \citep{Rivo11,Rivo13}. 
The models have depth varying density profiles, resulting from compression and temperature 
variations. They are constrained to have the same mass, radius and mean moment of
inertia as Mercury.
The core of the models is made of iron and the light element sulfur
and we use equations of state to compute the density as a function of depth.
For each given core radius, the radius of the inner core is determined from the sulfur 
concentration, temperature and pressure inside the core, and from the iron-sulfur melting
temperature. 
In this study we only consider models which have an inner core,
however we note that models without an inner core are also compatible with observations.
Likewise the core, for the mantle we solve equations of state to calculate the density as a 
function of depth. However, since the pressure increase with depth inside the mantle is small 
and since phase transitions to denser mineral phases are unlikely, we here neglect the small 
increase in density with depth and use a uniform density for the mantle (the average value of 
the depth dependent mantle density profile).
Therefore, and as a consequence of mass conservation the average mantle density for models of 
the same mantle composition vary with mantle temperature, core radius, and inner core radius 
(see Table 3).
Unlike for the mantle, compressibility effects in the core are significant and
cannot be neglected.
For the interior models we use 5 plausible mantle mineralogies that have been deduced 
from considerations about Mercury's formation and from the chemical composition of its surface 
\citep{Verh09,Rivo09}. The temperature inside the core is assumed 
adiabatic and for the mantle we use a cold ($T_{cmb}$=1850K) and a hot temperature 
($T_{cmb}$=2000K) profile that have been obtained from independent studies about the 
thermal evolution of Mercury (see the references in \citealt{Rivo09}).
These two values are representative of the lower and upper temperature range for Mercury
from thermal evolution. Many other mantle mineralogies and temperature profiles are possible, 
in particular profiles with significantly lower temperatures \citep{Grot11}.
Therefore it has to be kept in mind that the interior models that we use only represent a
subset of possible interior models.
Among the different combinations of mineralogies and temperature profiles, 9
different classes of interior models fulfill the constraints on mass, radius and
mean moment of inertia for some inner core radii (see Table 3). Interior models with a cold
temperature profile fulfill the constraints for a larger range of inner core
radii than interior models with a hot temperature profile.
As a consequence of global mass conservation and due to the fact that the solid inner 
core is denser than the liquid outer core, the radius of the core decreases with increasing 
inner core radius.
The hot models represent planets in which there is almost no sulfur in the core and in 
which the core is therefore denser.
\begin{center}
\begin{small}
\hspace{-1.2cm}
\begin{tabular}{|c|c|c c|c c|c c|}
\hline
 Model & CMB & \multicolumn{2}{c|}{$\rho_{m}$ (kg/m$^3$)} 
& \multicolumn{2}{c|}{$r_{icb}$ (km)}& \multicolumn{2}{c|}{$r_{cmb}$ (km)} \\
 name &temper. (K) &min & max &min & max&min & max \\
\hline
\multirow{2}{*}{FC} & 1850 & 3336 & 3395 &1 & 1810& 1820 & 1976 \\
 & 2000 & 3343 & 3355 &1 & 1830& 1837 & 1876 \\
\hline
\multirow{2}{*}{MA} & 1850 & 3275 & 3298 &1 & 1820& 1838 & 1983 \\
 & 2000 & 3281 & 3285 &1 & 1840& 1850 & 1883 \\
\hline
\multirow{2}{*}{TS} & 1850 & 3314 & 3344 &1 & 1860& 1830 & 1978 \\
 & 2000 & 3309 & 3317 &1 & 1830& 1843 & 1880 \\
\hline
 \multirow{2}{*}{MC} & 1850 & 3224 & 3245 &1 & 1840& 1847 & 1988 \\
 & 2000 & 3230 & 3233 &1 & 1840& 1859 & 1880 \\
\hline
 EC & 1850 & 3093 & 3115 &1350 & 1850& 1870 & 2002 \\
\hline
\end{tabular}
\end{small}

Table 3: Summary of the 9 interior models classes. Crust thickness is 50km and crust density 
is 2900 kg/$m^3$ for all models. See \citet{Rivo09} and \citet{Rivo13}
for more details on these models.
\end{center}

The spherical interior model is then transformed to a bi-axial model with equatorial flattening.
We assume that Mercury has a rigid bi-axial mantle and a core that is in hydrostatic equilibrium 
with respect to the mantle.
We specify the geometrical flattening at the top and bottom of the mantle ($\beta_m$ and $\beta_{cmb}$), 
and we compute the flattening as a function of radius in the core in response to the flattened 
mantle, following the method described in the study of \citet{Dumb13}. 
Flattening versus depth in the core is calculated under the assumption of hydrostatic equilibrium. 
In the following sections, we consider that $\beta_{cmb} = \beta_m$, $\beta_{cmb} = \beta_m/2$ and 
$\beta_{cmb} = 2 \,\beta_m$. Once the ratio between $\beta_m$ and $\beta_{cmb}$ is 
chosen, the flattening of the whole planet is adjusted to match the observational constraints 
on $C_{22}=(B-A)/(4 \,M R^2)$ obtained from measurements by MESSENGER \citep{Smit12}. 
We use the mean value of $C_{22}$ without assuming an error for all calculations.
From the resulting core flattening with depth, the coupling strength's 
$K$, $K_m$ and $K_{ic}$ can be computed \citep{Dumb13}.
With increasing inner core radius, the parameters $K_{ic}$ and $K$ increase while $K_m$ decreases.
The difference between the parameter values for the 3 flattening assumptions is 
less than 2\% for $K_m$, between 14\% and 30\% for $K_{ic}$, and up to more than 50\% for $K$.

\section{Results}
\subsection{Periods of the eigenmodes and resonance}
\label{sec_reso}
\begin{figure}[!ht]
\centering
\includegraphics[height=7cm]{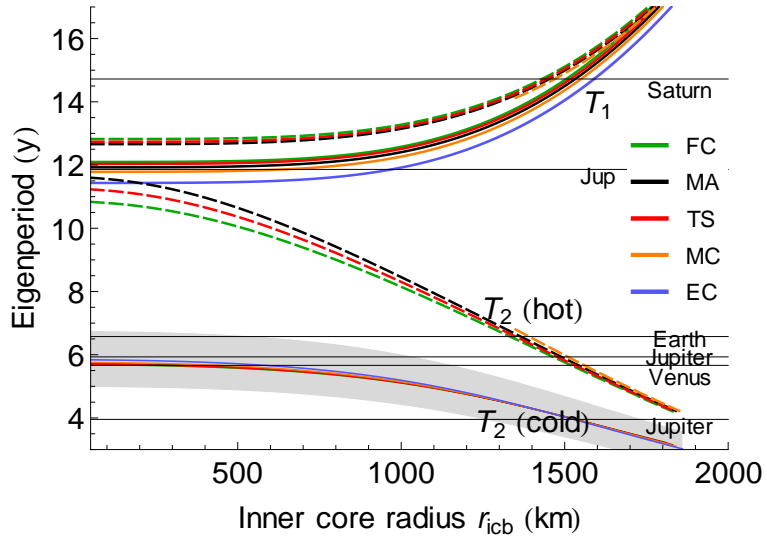} 
\caption{Periods of the two eigenmodes ($T_1$, $T_2$) 
as a function of the inner core size. 
The different curves represent the 9 different interior models classes.
The dashed curves represent the 4 models with the hot temperature profile.
The light gray areas show the spread of the free libration periods resulting 
from the flattening assumption, for the TS class of interior model. 
In this gray area, the flattening assumptions varies from $\beta_{cmb} = 2 \beta_m$ 
(lower limit) to $\beta_{cmb} = \beta_m/2$ (upper limit). 
The gray area is almost invisible around $T_1$.
The horizontal black lines show the 6 main periods of the planetary perturbations. 
The orange dashed curve does not extend below 1350km because for this particular mantle 
mineralogy (MC) and the hot temperature profile, it is not possible to find interior models 
with a small inner core that fulfill the mass and radius constraints.}
\label{fig_periods}
\end{figure}
The first free mode $T_1$ (from Eq.~(\ref{eq_om1})) represents the free libration of the combined mantle and 
inner core. Its period increases with inner core size (Fig.~\ref{fig_periods}). 
Without an inner core, this mode reduces to the mantle free libration period $2 \pi/\omega_m$ (Eq.~(\ref{eq_omm})) 
and its period is between $11.4$ and $13.4$y. 
For $\beta_{cmb} = \beta_m$, the period $T_1$ is between $11.4$ and $17.7$y,
approaching the value of about 18y for an almost fully solid Mercury.
It stays very close to $2 \pi/\omega_m$ for small inner cores but significantly differs from the 
mantle libration period for inner cores larger than about 1000 km. 
The effect of the flattening assumption on the first free period is very small: for $\beta_{cmb} = \beta_m/2$, 
$T_1$ varies between $11.4$ and $18$y and between $11.4$ and $17.6$y for $\beta_{cmb} = 2 \beta_m$. 
We plot in Fig.~\ref{fig_periods} a gray area covered by the different 
assumptions for the flattening, but this area is almost invisible. 

The period of the second free mode $T_2$ (mantle-inner core gravitational 
mode) covers a very large range,
depending on the mantle mineralogy, the temperature profile and the flattening assumption. 
Without mantle-inner core coupling, this mode does not exist. 
For $\beta_{cmb} = \beta_m$, $T_2$ is between $3$ and $5.8$y for interior 
models with a cold temperature profile and between about $4.1$ and $11.6$y for interior 
models with a hot temperature profile. 
In contrast to $T_1$, $T_2$ is significantly affected by the flattening assumption.
The free period $T_2$ varies between 3.6 and 12.8y for $\beta_{cmb} = \beta_m/2$ and between 2.5 
and 9.4y for $\beta_{cmb} = 2 \beta_m$. Hence, $T_2$ increases with decreasing CMB flattening. 
In Fig.~\ref{fig_periods}, the light gray area shows the large spread in the free libration 
periods resulting from the different flattening assumptions, for one class of interior models.

The period $T_2$ decreases with inner core size. This is because $T_2$ depends mainly on 
the strength of the mantle-inner core gravitational coupling (constant $K$), which depends 
primarily on the density contrast between the fluid and solid core at the ICB. 
In order to satisfy the constraints on Mercury's mass, models with large (pure Fe) inner 
core must have a larger weight fraction of sulfur in the fluid core and thus a larger 
density contrast at the ICB: the larger $K$ then leads to a shorter $T_2$ period. 
As it can be seen on Fig.~\ref{fig_periods}, different interior density models result in vastly 
different $T_2$ periods. There is a clear separation between the hot and cold mantle models; 
for a given inner core size, hot models have a much lower sulfur fraction in the fluid core, 
and thus have a smaller $K$ and a longer $T_2$ period. 

The hot and cold temperature are plausible temperature profiles for the mantle of Mercury; 
any intermediary temperature profile would produce a $T_2$ curve that lies in between that 
of the hot and cold models of Fig.~1.
However, as other temperature profiles are possible, the range of variation of $T_2$ 
can be appreciably larger.
Lastly, we note that the determination of $T_2$ crucially depends on the radial variations 
in density in the core \citep{Dumb13}. Taking instead the density of the fluid 
and solid cores to be uniform and equal to their mean values, the range of $T_2$ values on 
Fig.~\ref{fig_periods} would be much narrower, from 2.8y to 5.2y.

When a planetary period is near one of the free periods, the libration amplitude at that frequency is 
resonantly amplified. 
Whether a large amplification by resonance occurs (i.e. intersection of $T_1$ or $T_2$ with a planetary 
orbit period, see the horizontal lines in Fig.~\ref{fig_periods}) 
depends on the interior structure of Mercury, and the equatorial flattening with depth.
For an inner core radius smaller than about 1000km, the period $T_1$ is very close 
to the orbital perturbation by Jupiter at $11.86$y. A large inner core moves the free 
libration period $T_1$ further away from the $11.86$y perturbation period, and would result 
in a smaller long-period forced libration amplitude. 
However, the free period $T_1$ could be very close to the orbital perturbation by Saturn ($14.7$y) 
if the inner core radius were approximately $1500-1600$ km. The free mode period $T_2$ may be resonant 
with the $3.95$y, $5.66$y, $5.9$y and the $6.57$y orbital perturbations.
For hot interior models, the free period $T_2$ may be close to the $11.86$y perturbation 
period for small inner cores.

\subsection{Amplitude of the long-period forced librations}

\begin{figure}[!ht]
%\centering
\includegraphics[height=7.cm]{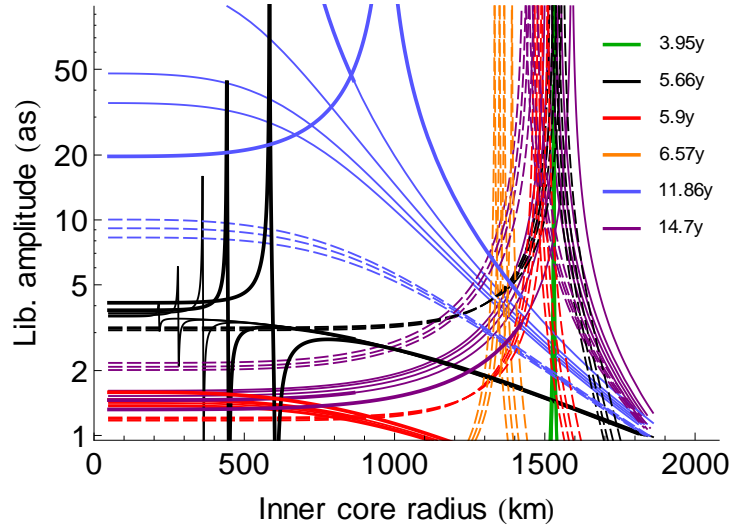}
\caption{
Absolute value of the amplitudes of the long-period forced variations of the rotation angle $\psi_m$ 
as a function of the inner core size for 9 different classes of interior models,
using damping parameters of $k_m=k_{ic}=10^{-5}$/y and no viscous inner core deformation 
($\tau \rightarrow \infty$).
The different colors show the 6 main perturbing frequencies.
We use dashed lines for the 4 classes of hot interior models.
The outer core flattening is $\beta_{cmb}=\beta_m$.
The thicker parts of the curves show the interior models for which the amplitude of the 
88d libration falls within the 1-sigma interval of \citet{Marg12}.
The amplitude of the 88d libration is about 38.5 as.}
\label{fig_amplit}
\end{figure}
Since the forcing frequencies are known from the orbital theory (Table 1), we
can compute the librations of the mantle and of the inner core for each of these
frequencies. 
The amplitudes computed in this section included viscous coupling at both the
CMB and the ICB ($k_m=k_{ic}=10^{-5}$/y), no electromagnetic coupling and
assumed a rigid inner core ($\tau \rightarrow \infty$). 
The level of amplification for each of the long-period forced librations differs for
different interior models of Mercury. 
This is because each of these models have different coupling parameters $K$,
$K_m$ and $K_{ic}$, which affects the frequencies $\omega_{1,\,2,\,m,\,g,\,ic}$
on which the amplitude depends (see equation \ref{eq_psimAn}). 
Fig.~\ref{fig_amplit} shows the absolute value of the 6 largest amplitudes of 
the long-period forced librations of the mantle $\psi_m$ as a function of the inner core 
size for the case $\beta_m=\beta_{cmb}$. 
A large amplification by resonance occurs when, for a specific inner core
radius, the period of one of the free modes is close to the period of an orbital
perturbation. 
The main long-period forced libration is the one due to Jupiter's perturbation on Mercury's
orbit at 11.86y period (blue curves in Fig.~\ref{fig_amplit}). As in the case
without inner core mantle coupling (see figure 5 of \citealt{Yseb10}), 
a large amplification of this libration occurs for inner cores that are not too large.
For an inner core radius smaller than 1200 km, the effect of the inner core on the eigenperiod $T_1$ 
is weak (Fig.~\ref{fig_periods}), and the amplitude of the 11.86y libration depends mainly on 
the ratio $(B-A)/C_m$ which determines the free mantle libration period (see Eq.~(\ref{eq_omm})). 
For hot interior models with small inner cores, the second free period $T_2$ can also be close to the 
11.86y period. 
However because the resonance factor associated with $T_2$ ($\omega_m^2-N$ coefficient 
in Eq.~(\ref{eq_psimAnDec})) is typically about 10 times smaller than the 
resonance factor associated with $T_1$, the amplification in that case is much smaller.

Long-period forced librations at five other frequencies may also have an amplitude larger than a few arcsec: 
at 3.95y, 5.66y, 5.9y, 6.57y and 14.7y periods. If Mercury does not have an inner core, the 
amplitude of these librations are not amplified and always remain below 5 as. 
Since the difference between the free periods and these forcing periods changes faster
with changing inner core radius than for the 11.86y libration (see Fig.~\ref{fig_periods})
and since the amplitude of the planetary perturbation $\lambda_i$ is small,
these resonances occur over a narrower range of inner core radii.

As the inner core radius increases from 50 to 600 km, for the cold interior models, the 5.66y
libration amplitude (black curves) varies from minus a few arcsec to $-\infty$ at the point 
when $T_2$ is precisely equal to 5.66y. Once $T_2$ crosses 5.66y, the resonance factor 
changes sign (because of $\omega_2^2 - \omega_i^2$ in Eq.~(\ref{eq_psimAnDec})) and the 
mantle libration amplitude jumps to $+\infty$. A further increase in inner core radius 
takes $T_2$ further away from the resonance and the mantle libration decreases back to 
minus a few arcsec. The quantity plotted in Fig~\ref{fig_amplit} is the absolute value 
of this amplitude, so it is remains positive. However, note that the mantle libration 
amplitude passes through zero for a specific inner core radius. This marks the location 
where the planetary forcing on the mantle is equal and opposite to the gravitational torque 
by the inner core. This does not occur precisely at $T_2$ = 5.66y, but when $T_2$ is slightly 
offset (and smaller). This effect is more visible for the 5.66y period than 
for other orbital periods because it has the largest orbital perturbation (see Table 1). 
For other orbital frequencies, since the orbital perturbations $\lambda_i$ are much smaller, 
the amplitudes far from the resonant frequencies are very small (close to zero) and 
the change of sign happens for libration amplitude smaller than 1 as, outside the graph limits.

In order to get the flattening of the mantle, we use the mean value of $C_{22}$ from 
\citet{Smit12} without assuming an error.
A different choice of $C_{22}$ within its $3\sigma$ error bound largely
affects the amplitude and location of 
the resonance because $C_{22}$ determines the amplitude of the planetary flattening, and therefore 
it affects the values for $K$, $K_m$, and $K_{ic}$. 
Because the free period $T_2$ is very sensitive to the flattening assumption,
a small change in the core equatorial flattening induces a change in the position of the resonance 
and the level of amplification. The choice of the $\beta_{cmb}/\beta_m$ ratio has the strongest effect 
on the resonance of the 3.95y, 5.9y and 6.57y periods. 
For the other orbital periods, the choice of the interior model is more important. 
As Fig.~\ref{fig_amplit} illustrates, the amplitudes of the long-period forced librations cannot be predicted
with any reasonable precision because of their strong dependence on 
interior parameters of Mercury for which we do not have good constraints.
Nevertheless, some general observations can be made. For inner cores smaller than about 
1200 km, the 11.86y libration is expected to have a forced libration amplitude of at least about 10 as.
Its amplitude decreases to about 1 as for interior models with very large inner cores. 
Detection of this forced libration might thus be used to distinguish between very large inner core or not.
If a large amplitude of the 14.7y libration is detected with a precision of a few arcseconds, 
then it means that the inner core is large (radius between 1300 and 1600 km).
The amplitude of the 3.95y libration might be large only if the inner core is very large. 
However, this is a very narrow resonance.
Finally, a very narrow resonant amplification of the 3 other planetary frequencies is 
possible for large or small inner core radius depending on whether the temperature of the 
mantle is hot, mild or cold.

The interior models in Fig.~\ref{fig_amplit} cover a large range of $(B-A)/C_m$ values, 
and therefore a large range of 88d libration amplitudes.
However a recent fit of the 88d libration amplitude to the rotation data 
\citep{Marg12} gives a 1-sigma interval for this amplitude of $[36.9-40.1]$ as.
Models that agree with the 1 sigma interval of the 88d libration are represented by 
thick lines in the figure~\ref{fig_amplit}. 
If we extend the uncertainty interval to a 3-sigma interval, then the number of allowed
interior models is much larger: almost all the models in the figure are allowed.

\begin{figure}[!ht]
%\centering
\includegraphics[height=6.5cm]{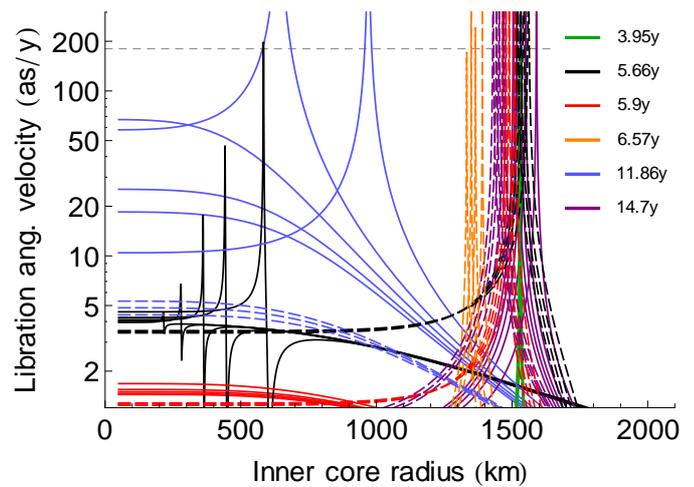} 
\caption{Absolute value of the amplitudes of Mercury's long-period librational angular 
velocity $\dot{\psi}_m$ for 9 different classes of interior models and for the 6 main 
orbital frequencies. The configuration is the same as 
for Fig.~\ref{fig_amplit}. The amplitude of the 88d angular velocity is of the order of 1000 as/y 
(out of the boundaries of the figure).
The horizontal dashed line shows the current uncertainty on the radar observations.}
\label{fig_amplit2}
\end{figure}

Figure~\ref{fig_amplit2} shows the amplitudes of the angular rotation velocity variations $\dot{\psi}_m$ 
(Eq.~\ref{eq_psimAn}) for the same 6 orbital perturbation frequencies and 
for the different interior models. 
Resonant amplifications are present at the same frequencies as the libration angle for specific 
interior models (see Fig.~\ref{fig_amplit}).
Their amplitude is usually smaller than 80 as/y, except for a very small range of interior models.

\subsection{Effect of the dissipation}
Calculations for different choices of $k_m$ and $k_{ic}$ than the 
nominal values $k_m=k_{ic}=10^{-5}$/y show that, provided $k_m$ and $k_{ic}$ are not larger than 
approximately $10^{-3}$/y, the only difference with the results 
presented in the previous section is a reduction of the peak amplitude of libration at the resonance periods. 
As discussed in section 1.5, the addition of EM coupling at the CMB should not alter the amplitude of $k_m$ 
significantly, whereas EM coupling at the ICB could lead to an amplitude of $k_{ic}$ of the order of $10^{-3}$/y 
for reasonable assumptions about the magnetic field strength. Therefore, except at the precise location of resonances, 
the results for the nominal case presented above
are not largely affected by the nature and amplitude of coupling at the CMB and ICB.
The maximal amplitude of the peak may be approximated by Eq.~(\ref{eq_psimimax}).
The damping parameter $k_m$ has a larger effect than the damping parameter $k_{ic}$
on the libration amplitude of the mantle and of the inner core.
This effect is also observed in the maximal amplitude for $\psi_m$. 
This can be understood from the expression (Eq.~(\ref{eq_psimimax})) of the maximal amplitude for $\psi_m$:
in the denominator, the coefficient in front of the damping parameter $k_m$ is much smaller than 
the one in front of $k_{ic}$.
Dissipation can significantly change the phase of the libration but 
since the electromagnetic and viscous couplings are usually smaller than the gravitational coupling, 
this only happens if a planetary perturbation frequency is very close to the free mode frequency. 
If the differences between the frequencies $\omega_1 - \omega_i$ and $\omega_2 - \omega_i$ 
are larger than about $0.002$ rad/y, the phase angle $\phi_{\psi_i}$ (Eq.~(\ref{eq_psiphase})) reduces 
to $\phi_{\lambda_i}$ or to $\phi_{\lambda_i} + 180 ^\circ$ (the value depends on 
the sign of the frequency differences) with an error of less than $0.2^\circ$
for the nominal value of the dissipation parameters. 
The widths of the resonance peaks depend mostly on the amplitude of the planetary 
perturbations $\lambda_i$ and only very slightly on the damping parameters. 

Calculations with different choices of the inner core viscous relaxation time $\tau$ show that, 
provided $\tau$ is larger than about 25 years, the main difference in the results is a 
reduction of the peak amplitude of 
librations at the resonance. 
Thus, except for inner core radii where the free period $T_1$ or $T_2$ almost matches a 
planetary forcing period, 
our results in the previous section are barely sensitive to the precise value of $\tau$. 
However, if $\tau$ is of the order of about 25y or less, then 
the amplitude of the long-period librations are reduced for a much broader range of inner core radii.
If $\tau$ is less than a few years, then the amplitudes of the long-period librations are almost 
reduced to the original orbital perturbation $\lambda_i$.
The phases of the long-period librations are also largely changed.

\subsection{Link with the observations}
The amplitude of the 88d forced libration (38.5 as, \citealt{Marg12}) 
has approximately the same order of magnitude as the long-period forced librations (see Fig.~\ref{fig_amplit}). 
This is not the case 
for the amplitudes of the angular rotation velocity variations $\dot{\psi}_m$ (see Fig.~\ref{fig_amplit2}):
the 88d amplitude is about 1000 as/y, 
about 10 to 100 times larger than the long-period amplitudes.
This is due to the temporal derivative that leads to more prominent short-period effects in the rotation 
rate than in the libration angle.
The long-period librations will be more difficult to detect in spin radar data 
than in orientation data.

These amplitudes may also be compared to the present uncertainty on the observations of the rotation state 
of Mercury. The average precision on the rotation rates obtained with Earth-based radar data 
\citep{Marg12} is of the order of 180 as/y. 
Fig.~\ref{fig_amplit2} shows that a velocity amplitude larger than 180 as/y only 
occurs under a very specific range of inner core radii (except in the case of the
11.86y libration period for which the signature may be detectable over a larger
range of inner core radii).
Given the current observational precision, it follows from the calculated libration velocities, 
that a large difference between models with and without inner core 
coupling can only be seen if Mercury happens to be close to a resonance.

The libration angle can also be measured directly.
A measurement of the libration amplitude was recently provided by \citet{Star12} 
using MESSENGER data with an uncertainty on the 88d libration amplitude of about 5 as.
The expected precision on the libration angle with the BepiColombo spacecraft is about 1 arcsecond 
\citep{Pfyf11}.
Fig.~\ref{fig_amplit} shows that libration signatures for a large fraction of interior models
would exceed a few arcseconds and would therefore be detectable.
Therefore, the influence of the inner core in the long-period forced librations has a better chance to be 
detected on the basis of future observations of the libration angle rather than the rotation rate.

In order to gain pertinent inferences about the interior structure of Mercury, one would need 
arcsecond precision observations of the libration angle over a time interval long enough to 
identify the planetary periods, i.e., 5-15y.
Clearly, predictions of the long-period forced librations from our model depend on many 
parameters that are not well known such as the inner core size, the interior density structure, etc. 
Different combination of these parameters can generate long-period librations of equivalent amplitudes. 
Nevertheless, despite the non-uniqueness, the constraint of matching predicted with observed long-period 
librations would reduce the possible parameter space and thus provide valuable information on Mercury's interior.

The 88d libration amplitude of the inner core $\gamma_{ic}$ is much smaller 
than the 88d libration amplitude of the mantle \citep{Vanh12}.
The amplitude of the 88d libration is between $0.5$ and $2.2$ as, depending on 
the interior model and the flattening assumption.
At semi-annual period, the inner core libration is below 0.2 as \citep{Vanh12}.
However, the long-period librations of the inner core may also be largely amplified,
for the same frequencies as the mantle libration
since the denominators in the equations for the libration amplitude are the same 
for the inner core and for the mantle (Eqs.~(\ref{eq_gamAn}) and (\ref{eq_gasAn})).
Although inner core librations could in principle be detected through their effect 
on the degree-two, order-two gravitational coefficient $C_{22}$, the signal is probably too
weak to be observed in the near future by either MESSENGER or BepiColombo.

\section{Fit of the radar data}

In this section, we investigate whether including the 
coupling between the mantle and inner core can improve the fit to the 
radar data of \citet{Marg12} and determine whether it is 
possible to constrain the size of the inner core from the measurements of the rotation rate.

Earth-based radar measurements of the instantaneous spin state of Mercury have been made 
at 35 different epochs between 2002 and 2012 \citep{Marg12}.
From the correlation of the radar echo signals at two Earth stations,
\citet{Marg12} obtained 35 instantaneous spin rate values.
Currently, the averaged uncertainty on these rotation rates is about 180 as/y. 
Since two polarizations have been recorded, two data points 
and the associated uncertainties are available at each epoch.\\
We fit the observed rotation rates with a rotation model for Mercury that includes
the short-period forced librations (annual, semi-annual, etc., see section \ref{sec88}) and the long-period 
forced librations.
The model also includes dissipation at the CMB and ICB (damping parameters $k_m=k_{ic}=10^{-5}$/y).
We assume that the two free librations have not been excited recently (e.g. by an internal forcing, 
\citealt{Koni13}) and have attenuated to a negligible amplitude. 
Adding a free libration would also complicate the analysis by introducing additional 
parameters to estimate and by a possible overlap with the 11.86y libration.
In order to compare the solution without an inner core with a situation with an inner core,
we use as an estimator of the goodness of the fit the reduced $\chi^2$ 
(sum of the squares of the differences between the model and the data expressed in units of the uncertainty, 
divided by the number of degrees of freedom). 
We seek a global minimum of the reduced $\chi^2$ for all the interior models.

\begin{figure}[!ht]
\centering
\includegraphics[height=5cm]{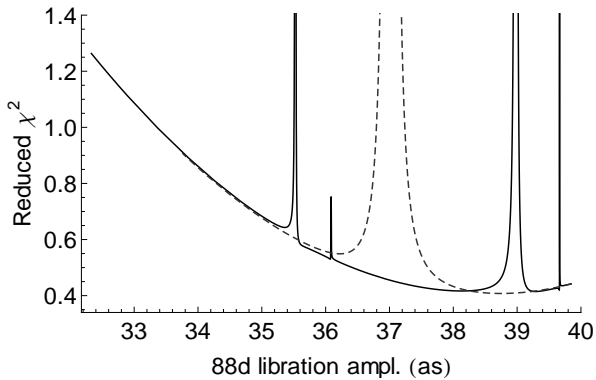} 
\caption{Reduced $\chi^2$ as a function of the annual libration amplitude for the EC interior model 
class with a cold temperature profile. 
The inner core radius increases from right to the left.
The assumption for the flattening is $\beta_{cmb}=\beta_m$.
The black (dashed) line shows the reduced $\chi^2$ using a rotation model with (without) inner core 
coupling.}
\label{fig_chi2}
\end{figure}
Fig.~\ref{fig_chi2} shows that the main effect on the reduced $\chi^2$ is the 
88d libration amplitude. The reduced $\chi^2$ has a strong dependence on the 88d 
libration amplitude, both for interior models that take into account the mantle-inner core coupling 
and for those that do not.
Additionally we see that the long-period forced librations increases the reduced $\chi^2$ 
for very specific interior models.
To match the constraints on Mercury's mass, models with large inner core have smaller outer core 
radius; this leads to an increase in $C_m$ and thus a decrease in $(B-A)/C_m$ and in the 88d 
amplitude.
Therefore the interior models with small inner cores have 88d libration amplitude larger 
than 39 as in Fig.~4 while models with large inner cores have a 88d libration amplitude smaller
then 35 as.

In general, the inclusion of mantle-inner core coupling in the rotation model does not 
improve the best fit (Fig.~\ref{fig_chi2}).
The global minimal reduced $\chi^2$ is about $0.4$, which 
is about the same as for the situation without mantle-inner core coupling.
The effects of a small inner core (radius smaller than about 500km) on the rotation model 
are usually so small that the rotation models with and without mantle-inner core coupling 
produce almost the same $\chi^2$. 
However if a resonance between one of the free modes and one of the planetary 
perturbation periods occurs for some particular interior model, the amplitude 
of the long-period libration becomes very large and the fit is usually degraded. 
Thus, based on the reduced $\chi^2$ analysis, we conclude that the radar observations suggest that 
Mercury's librations do not contain large, resonantly amplified, long-period forced librations.
Other flattening assumptions change the position of the resonances, but 
do not globally decrease the reduced $\chi^2$.

\begin{figure}[!ht]
\centering
\includegraphics[height=5cm]{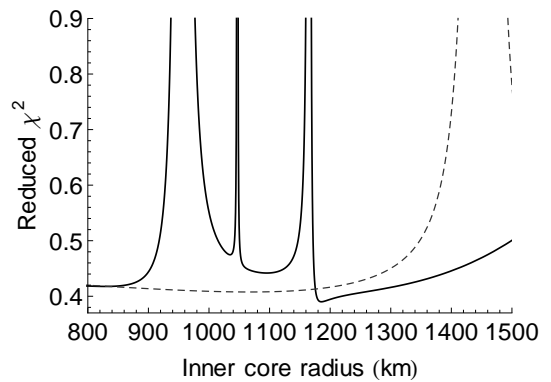} 
\caption{Reduced $\chi^2$ as a function of the inner core radius for the EC interior model 
and a cold temperature profile (see \citealt{Rivo09} for details). 
The assumption for the flattening is $\beta_{cmb}=\beta_m/2$ and $C_{22}=8.088 \times 10^{-6}$.
The black (dashed) line shows the reduced $\chi^2$ using a rotation model with (without) 
mantle-inner core coupling.
We see that a global minimum for the reduced $\chi^2$ is when the inner core radius is $1186$km,
indicating the best fit model.}
\label{fig_chi2GloMin}
\end{figure}
Some interior models have been found that have a reduced $\chi^2$ smaller than 0.4
when inner core coupling is included. For example, the reduced $\chi^2$ is $0.39$ 
(see Fig.~\ref{fig_chi2GloMin}) if $\beta_{cmb}=\beta_m$/2, $r_{icb}= 1186$ km and using a cold 
mantle with a low density (EC model of \citealt{Rivo09}).
This is due to a resonance between the 5.66y perturbation and the free period $T_2$. 
For some specific inner core radii close to 1186km, an amplitude of 23 as 
for the 5.66y libration allows for a better fit to the observations. However, the improvement in 
the fit is extremely sensitive to the precise amplification of the 5.66y libration: a small increase 
in the amplification leads to a large reduced $\chi^2$ and degrades the fit dramatically.

If we explore the confidence interval for $C_{22}$ and the other parameters of our model, 
it is also possible to find very sharp resonances that can, for very specific interior model 
and inner core radius, slightly decrease the reduced $\chi^2$.

The effect of the inner core can therefore provide a slightly better fit to the data for specific 
interior structure models. However, the improvement is only marginal and it cannot be reasonably 
argued that they provide a better match to the libration observations. 
The uncertainty on the available data is presently too large to be sensitive to the effect 
of mantle-inner core coupling and by extension, to the size of the inner core. 
Additionally, the time period covered by the available data is too short to
separate the different libration amplitudes.
Since a dramatic worsening of the fit may occur at resonances, we can only conclude that a very large amplification 
is most likely not present in Mercury's librations. Based on this, combinations of interior parameters that 
would yield a free mode period very close to one of the main long-period librations must be rejected.

\section{Conclusion}
We presented a theory to compute the amplitude and phase of the long-period
forced librations of Mercury. These long-period librations are caused by the planetary
perturbations on the orbit of Mercury. Our model takes into account the
internal coupling that occurs between the solid inner core, fluid outer core and
mantle. This includes gravitational and pressure torques between these layers,
as well as viscous and electromagnetic couplings at the CMB and ICB. Our model
also includes viscous deformation of the inner core. 

Previous studies that have investigated the long-period forced librations of Mercury
did not take into account the coupling dynamics of the inner core with the rest
of the planet. The inclusion of the inner core leads to two eigen modes
that have a large range of possible periods. Their precise periods depend on
the interior structure, including the size of the inner core. Large
amplification of a long-period forced libration occurs if its period is close to one of
the free mode periods. Our results show that for a large set of models, long-period forced librations
may have an amplitude well above 5 arcsec. 
For a small inner core radius, the first free period $T_1$ is close to the Jupiter 
perturbation period at $11.86$y while for a large inner core, it can be closer to Saturn orbital 
perturbation at $14.7$y. The second free period may be resonant with several orbital 
perturbations: $3.95$y, $5.66$y, $5.9$y, $6.57$y and even the $11.86$y perturbation.
Except very close to resonance periods, the long-period forced librations are not
sensitive to the strength and nature of the coupling at the CMB and ICB 
if the damping parameters $k_m$ and $k_{ic}$ are smaller than $10^{-3}$/y. Our
results are also insensitive to viscous deformation within the inner core, 
except for the maximal amplitude of the resonances peaks,
provided they occur on a characteristic timescale longer than 25y. 

We also tested whether the inclusion of an inner core can provide a better fit
to the Earth based radar observations of Mercury's rotation rate. Because the
long-period forced librations have amplitudes that are below the current uncertainty on
these observations and because the time interval they cover is short compared to
the main planetary forcing periods, the data are not very sensitive to these long-period
forced librations. Therefore, it is not possible to detect the dynamical influence of
the inner core on Mercury's forced librations on the basis of the currently available
observations, other than the fact that very large forced libration amplitude at
planetary periods are not suggested by observations.

The amplitude of the long-period forced librations predicted by our model are sufficiently
large to be measurable by spacecraft observations of the
libration angle of Mercury. This is provided the libration angle is observed
for a sufficiently long time window and with a precision of a few arcseconds. 
Depending on their precision, future measurements of long-period forced librations might then allow us to place
useful constraints on the interior structure of Mercury. As we have
shown in our study, the combination of parameters (inner core size, interior
profiles of flattening and density, etc) that can generate a given 
libration is non-unique. Nevertheless, matching predicted and observed long-period forced librations 
would reduce the possible parameter space and thus provide
valuable information on Mercury's interior.

\section*{Acknowledgments}
This work was financially supported by the Belgian PRODEX program managed by the European Space Agency 
in collaboration with the Belgian Federal Science Policy Office.
MD is currently supported by a Discovery grant from NSERC/CRSNG.
We acknowledge J.L. Margot, S. Peale and an anonymous reviewer for their useful comments which helped us 
to improve the paper.

%\bibliography{innercore_articlebib}
\bibliographystyle{/usr/share/texmf/tex/latex/elsarticle/elsarticle-harv}

\end{document}